\documentclass[letterpaper]{article} 
\usepackage[]{aaai24}  
\usepackage{times}  
\usepackage{helvet}  
\usepackage{courier}  
\usepackage[hyphens]{url}  
\usepackage{graphicx} 

\usepackage{booktabs}
\usepackage{multirow}
\usepackage{multicol}
\usepackage{tabularx}
\usepackage{amsmath}
\usepackage{algorithm}
\usepackage{algorithmic}
\usepackage{subcaption}
\usepackage{tikz}
\usepackage{pgfplots}

\usepackage{natbib}  
\usepackage{caption} 
\usepackage{algorithm}
\usepackage{algorithmic}

\usepackage{newfloat}
\usepackage{listings}
\DeclareCaptionStyle{ruled}{labelfont=normalfont,labelsep=colon,strut=off} 
\floatstyle{ruled}
\newfloat{listing}{tb}{lst}{}
\floatname{listing}{Listing}
\title{N-Gram Unsupervised Compoundation and Feature Injection for Better Symbolic Music Understanding}
\author{Jinhao Tian$^{1,\dag}$, Zuchao Li$^{1,}$\thanks{$\ $  Corresponding author. $^\dag$ Equal contribution. This work was supported by the National Natural Science Foundation of China (No. 62306216), the Natural Science Foundation of Hubei Province of China (No. 2023AFB816), the Fundamental Research Funds for the Central Universities (No. 2042023kf0133), National Natural Science Foundation of China [No. 72074171] [No. 72374161]. }, Jiajia Li$^{2,3,\dag}$, and Ping Wang$^{2,3, *}$\\
$^{1}$National Engineering Research Center for Multimedia Software, \\
School of Computer Science, Wuhan University, Wuhan, 430072, P. R. China \\
$^{2}$Center for the Studies of Information Resources, Wuhan University, Wuhan 430072, China\\
$^{3}$School of Information Management, Wuhan University, Wuhan 430072, China
{\tt \{jinhaotian,cantata,zcli-charlie,wangping\}@whu.edu.cn}\\
}

\usepackage{bibentry}
\usepackage{url}

\begin{document}

\maketitle

\begin{abstract}
The first step to apply deep learning techniques for symbolic music understanding is to transform musical pieces (mainly in MIDI format) into sequences of predefined tokens like note pitch, note velocity, and chords. Subsequently, the sequences are fed into a neural sequence model to accomplish specific tasks.
Music sequences exhibit strong correlations between adjacent elements, making them prime candidates for N-gram techniques from Natural Language Processing (NLP). Consider classical piano music: specific melodies might recur throughout a piece, with subtle variations each time.
In this paper, we propose a novel method, NG-Midiformer, for understanding symbolic music sequences that leverages the N-gram approach. Our method involves first processing music pieces into word-like sequences with our proposed unsupervised compoundation, followed by using our N-gram Transformer encoder, which can effectively incorporate N-gram information to enhance the primary encoder part for better understanding of music sequences.
The pre-training process on large-scale music datasets enables the model to thoroughly learn the N-gram information contained within music sequences, and subsequently apply this information for making inferences during the fine-tuning stage.
Experiment on various datasets demonstrate the effectiveness of our method and achieved state-of-the-art performance on a series of music understanding downstream tasks. The code and model weights will be released at \url{https://github.com/CinqueOrigin/NG-Midiformer}.
\end{abstract}

\section{Introduction}

Symbolic Music Understanding, distinct from audio-based understanding~\cite{nam2018deep}, involves the computational analysis and interpretation of symbolic music sequences. This understanding aids tasks like music generation~\cite{briot2017deep} and Music Information Retrieval (MIR)\cite{casey2008content}. The process starts by converting music pieces into sequences of predefined tokens, representing musical events such as note pitch and tempo. These sequences are then processed using neural networks, notably the Transformer\cite{vaswani2017attention}. Music tokenization methods fall into two categories: direct conversion methods like MIDI-LIKE~\cite{oore2020time} and REMI~\cite{huang2020pop}, and methods employing expansion and compression techniques~\cite{li2021text}, such as Compound Word (CP)\cite{hsiao2021compound} and OctupleMIDI\cite{zeng2021musicbert}.

\begin{figure}[h]
    \centering
    \includegraphics[width=0.45\textwidth]{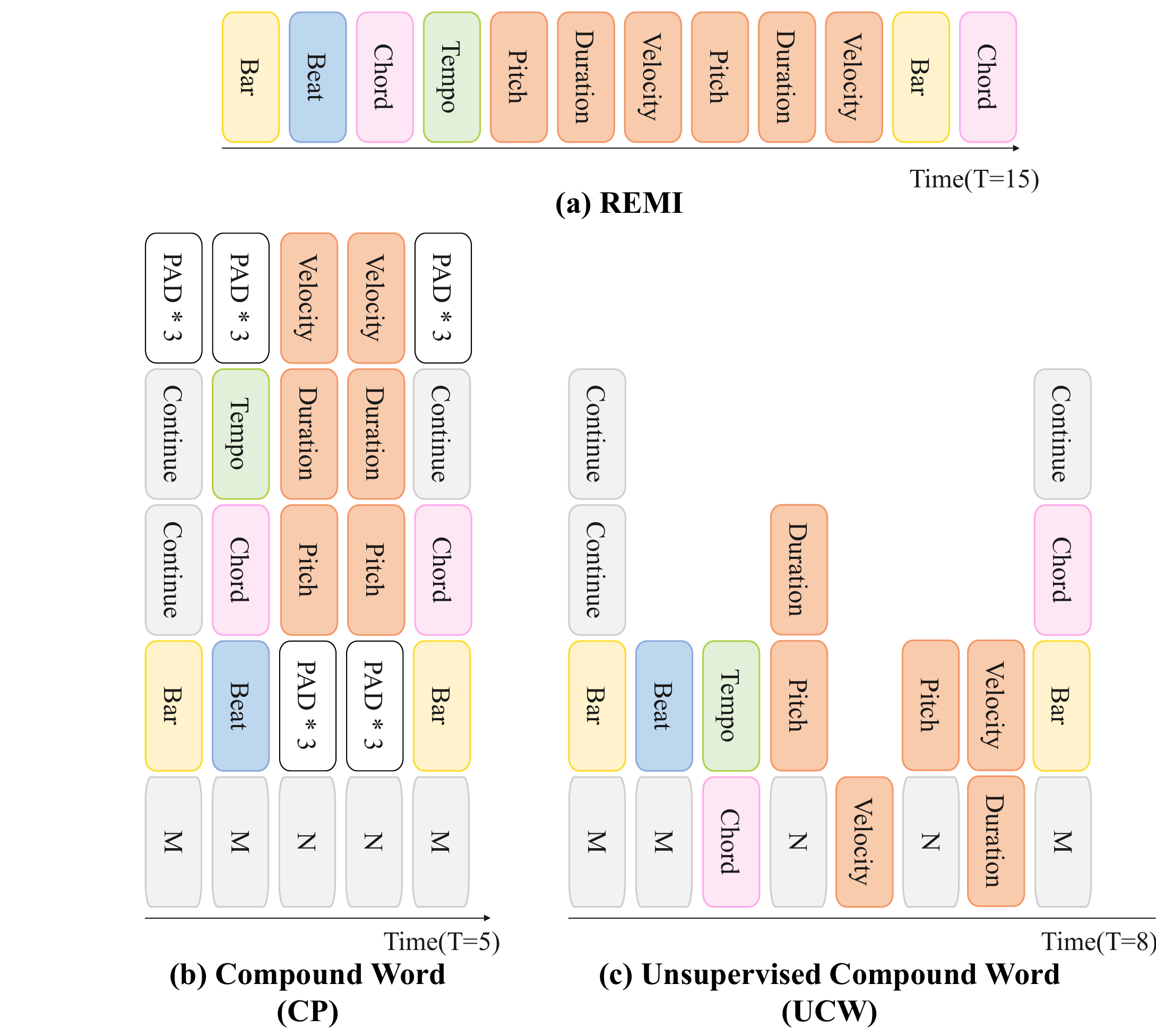}
    \caption{An example illustrating the relationships and distinctions among REMI, CP, and UCW.
    Here, M denotes the family ``metric", and N denotes the family ``note".}
    \label{fig:token}
\end{figure}

Direct conversion methods like MIDI-LIKE and REMI maintain music event atomicity but tend to produce extended sequences, weakening token dependencies. On the other hand, techniques like CP and OctupleMIDI, which utilize expansion and compression, yield shorter sequences rich in contextual information. However, they risk missing critical short-range dependencies between adjacent elements and often introduce redundant ``[PAD]" tokens or replicate neighboring musical tokens, diminishing the sequence's informational value. It's pivotal to note that music events often exhibit co-occurrence regularities, with certain events frequently appearing together, mirroring the music's semantic depth. This regularity necessitates a tokenization method that resonates with the unique characteristics of music events. Interestingly, music event atoms exhibit combinatory traits akin to characters in natural language. Motivated by this parallel, we propose an unsupervised compoundation method based on frequency. By analyzing the co-occurrence patterns of music events, we craft sequences using a frequency-driven vocabulary, ensuring a more concise sequence length than REMI while preserving the inherent correlations of compound words.

The success of sequence neural networks in NLP across diverse languages has provided valuable insights for symbolic music understanding~\cite{vaswani2017attention,devlin2018bert}. Transferring these models to symbolic music has proven effective in music generation and understanding \cite{hsiao2021compound,chou2021midibert,zeng2021musicbert}. However, existing models seldom undertake further processing to handle the strong correlation between consecutive adjacent musical elements, a feature that distinguishes music from language. Some methods even inadvertently reduce this correlation. Music often showcases repeating pattern both locally and across sections. The N-gram model, designed to capture local patterns, aligns well with this correlated nature of music and can be instrumental in identifying and leveraging these repetitions for enhanced the representation of symbolic music. Furthermore, music's hierarchical structure, ranging from individual notes to entire compositions, can be captured by N-grams, especially at smaller scales. Despite this alignment, few current models utilize N-grams in symbolic music understanding. We introduce an N-Gram Transformer encoding that capitalizes on the strong correlation of neighboring music events, improving music sequence understanding.

Unsupervised compoundation, a novel music tokenization technique, extracts N-gram information between neighboring music events. It synergizes ideas from previously mentioned tokenization methods. Initially, continuous music elements from the same ``family" are grouped into a ``word", as defined in Compound Word \cite{hsiao2021compound}. Here, a ``family" refers to a specific category of musical event, like pitch, duration, and velocity of a note, all under the ``note" family. Other families include metric and track, each representing distinct musical categories.
Subsequently, Byte Pair Encoding (BPE), an unsupervised method, segments these ``words". In the REMI sequence context, this means frequently occurring neighboring music events from the same family merge into a token, with each event akin to a ``character" in a ``word". Notably, our method omits the [PAD] token present in CP. Figure \ref{fig:token} delineates the relationships among REMI, CP, and our Unsupervised Compound Word (UCW). This technique not only shortens sequences compared to REMI but also sidesteps the excessive, often meaningless, [PAD] tokens seen in CP.

After UCW sequence construction, an encoder is employed to extract N-gram features, enhancing the model's symbolic music understanding via N-gram tokens. Our model, NG-Midiformer, incorporates a primary encoder for the input music sequence and an N-gram Transformer encoder to harness N-gram sequence information. This encoder, rooted in the Transformer architecture, is tailored to exploit N-gram information. To operationalize this, we commence by deriving N-grams from the given corpus and obtaining their frequencies, thereby constructing an N-gram vocabulary. For every input music sequence, we extract the relevant N-grams and pinpoint their positions within this established vocabulary. This comprehensive approach aligns with the intricate nature of symbolic music, and can capitalizes on the global structure inherent in musical compositions.

Our method's distinctiveness emerges from its integration mechanism. At every primary encoder layer, the output from the N-gram encoder is integrated, enriching the representation with N-gram contextual information. Specifically, for each N-gram, its hidden layer output is multiplied by its frequency, normalized, and then added to the primary encoder's hidden layer output, ensuring a comprehensive musical context understanding.

We pre-train NG-Midiformer on a large-scale, unlabeled symbolic music dataset, subsequently assessing its efficacy across six downstream tasks: composer, emotion, genre, and dance classification, as well as velocity prediction and melody classification. Our NG-Midiformer outperformed prior state-of-the-art models across these tasks, showcasing its superiority on datasets like Pianist8~\cite{joann8512_2021_5089279}, EMOPIA~\cite{EMOPIA}, GTZAN~\cite{sturm2013gtzan}, Nottingham~\cite{Nottingham}, and POP909~\cite{wang2020pop909}. This establishes our method as a robust new benchmark in symbolic music understanding.

\section{Related Work}

\subsection{Music Tokenization}

Musical compositions, akin to natural language, have ``grammatical" and ``semantic" structures, enabling their representation as structured sequences \cite{patel2003language,li2006toward}. While the pianoroll method encodes music into matrices differentiating pitch and time, it often requires fixed-length music event processing, reducing efficiency. Currently, MIDI files have emerged as a dominant symbolic representation, being lightweight and capturing essential musical elements.

Methods like MIDI-LIKE \cite{oore2020time} and REMI \cite{huang2020pop} provide detailed musical information but result in extended sequences. Compound Word (CP) \cite{hsiao2021compound}, a refinement of REMI, groups related music elements, shortening sequences. OctupleMIDI \cite{zeng2021musicbert}, an evolution of REMI and CP, offers structured encoding for diverse music types, being more concise than CP. However, both CP and OctupleMIDI introduce substantial filler information in tokens and employ multiple independent embedding layers, potentially weakening the relationship modeling between music events.

To tackle these challenges, we present an unsupervised compoundation method. This new tokenization strategy consolidates frequent adjacent music elements into one token, streamlining embedding processes and boosting efficiency. Using N-Gram data from music event groups, our approach delves deeper into music's semantic layers.

\subsection{Symbolic Music Understanding}

The development of symbolic music understanding is closely related to advancements in NLP techniques \cite{jackendoff2009parallels}. This is because both music and natural language can be represented as sets of symbols with certain structures and rules. Based on Word2Vec \cite{mikolov2013efficient,mikolov2013distributed} in NLP, researchers have grouped different musical notes together and treated them as a single unit or ``word". They then trained deep learning models on the resulting sequences \cite{hirai2019melody2vec}.

The emergence of models like Transformers \cite{vaswani2017attention}, BERT \cite{devlin2018bert}, GPT-3 \cite{floridi2020gpt} and so on~\cite{zhang2020semantics} have not only revolutionized NLP but also enriched symbolic music understanding. This is evident in the enhancements seen in models like Transformers-XL \cite{dai2019transformer} and CP Transformers \cite{hsiao2021compound}. MIDI-Bert \cite{chou2021midibert} and MusicBert~\cite{zeng2021musicbert}, both large-scale music pre-training models, exemplify the successful adaptation of NLP techniques, particularly BERT's architecture and Roberta's structure \cite{liu2019roberta}, to symbolic music. These models simply transfer the methods of NLP to symbolic music sequences and make some modifications.

However, music's inherent co-occurrence patterns, such as consistent events from certain chords, necessitate a tailored approach. Meanwhile, N-grams possess the capacity to capture such co-occurrence patterns~\cite{brown-etal-1992-class,sari-etal-2017-continuous,shafiq2008embedded}. To this end, we champion an encoding structure accentuating N-Gram features to enhance the representation of symbolic music. While NLP's progress is commendable, direct N-Gram technique adoption remains sparse. Inspired by ZEN~\cite{diao2019zen}, our proposed N-Gram Transformer structure discerns N-Gram relationships in symbolic music sequences, bridging this gap.

\section{NG-Midiformer}

Symbolic music events have highly pronounced local dependencies, and N-grams are particularly suitable for sequences with strong local dependencies. In this section, we propose a NG-Midiformer model, a powerful architecture for processing symbolic music sequences. The core concept underlying our model is N-gram, which consists of two key aspects: transforming music into appropriate tokens according to N-gram within the events family, and using N-gram between tokens within the sequence to enhance the understanding ability of the model.

\subsection{Music Tokenization}

To achieve symbolic music understanding using deep learning models like the Transformer, musical pieces must first be converted into symbolic element sequences. Given a musical piece $X$ in MIDI format, a mapping function $f$ transforms it into a sequence $S$ using predefined music tokens $e$ from vocabulary $V^{raw}$:
\begin{equation}
S = f(X) = \{e_1,e_2,...,e_N\}
\label{makeseq}
\end{equation}
where $N$ represents the sequence length, and $e_i$ is the $i$-th predefined musical token.

However, the self-attention mechanism in the Transformer struggles with longer sequences, such as those produced by the REMI method. To mitigate this, the CP method aggregates music events from the same family into compound words.

Given a REMI sequence \( \mathcal{R} \), the CP representation is constructed as follows.
Each \( CP_j \) is represented as:
\begin{equation}
    CP_j = [e^{cp}_{j,1}, e^{cp}_{j,2}, \dots, e^{cp}_{j,K}]
\end{equation}
where $K$ represents a fixed number of music element types in a compound word.

As for the definition of ``family", We partition these $K$ types into several non intersecting families with each representing a specific music family, such as note and metric. For instance, if $\mathcal{K}$ represents the type set corresponding to K and is partitioned into $c$ families, then the relationship among these families adheres to the following:
\begin{equation}
    \begin{aligned}
        \forall i \neq j, \mathcal{K}_i &\cap \mathcal{K}_j = \emptyset \\
        \bigcup_{i=1}^{c} \mathcal{K}_i &= \mathcal{K}
    \end{aligned}
\end{equation}
where $\mathcal{K}_i$ is the i-th family.

For each $CP_j$, identify a continuous segment in $\mathcal{R}$ where all elements belong to the same family. Let's denote this segment as $\mathcal{S}_j$.
Then the elements \( e^{cp}_{j,t} \) in \( CP_j \) are then defined by:
\begin{equation}
\setlength{\abovedisplayskip}{5pt}
\setlength{\belowdisplayskip}{5pt}
e^{cp}_{j,t} =
\begin{cases}
\mathcal{R}_{\mathcal{F}(j,t)}, & \text{if } 
\begin{aligned}
    &t \text{ matches the family of } \mathcal{S}j  \\
    & \text{ and }\mathcal{R}_{\mathcal{F}(j,t)} \text{ is in } \mathcal{S}_j 
\end{aligned} \\
\textrm{[PAD]}, & \text{otherwise}
\end{cases}
\end{equation}
where $\mathcal{F}(j,t)$ is a mapping function that converts the token index $j$ and inner token index $t$ in a CP into the index in the original REMI sequence. 

This method may lose crucial short-range dependency information between neighbouring elements as it might not capture the full relationships within a CP token.

To enhance existing music tokenization methods, we introduce an unsupervised compoundation approach tailored for music event representation. This method employs variable-length token design, grouping frequent music events within families by co-occurrence frequency. Unlike CP, we utilize a unified embedding based on segmented compound subwords, offering a closer semantic representation of music.
To construct the sequence, we adopt the concept of creating sub-word units from the domain of NLP. Several methods are available for creating sub-word units in NLP, such as SentencePiece \cite{kudo2018sentencepiece}, WordPiece \cite{wu2016google}, Byte Pair Encoding (BPE) \cite{sennrich2015neural} and so on~\cite{zhang2019effective}. The first two methods are similar to BPE with minor variations in implementation techniques. Therefore, we leverage the unsupervised BPE method to construct our UCW sequences.
Combining the ideas of REMI and CP, we construct music elements in the same way as REMI. Then, neighbouring elements belonging to the same ``family" in REMI are merged together as a family token according to the concept of ``family" in CP. Since we only compound the real music events according to co-occurrence frequency, [PAD] token presented in CP is not needed. Prior to constructing UCW, it is necessary to get the corresponding REMI vocabulary $\mathcal{V}_{REMI}$ and set the size of the UCW vocabulary $V_{UCW}$ artificially. This way, the UCW token can be represented as:
\begin{equation}
\setlength{\abovedisplayskip}{5pt}
\setlength{\belowdisplayskip}{5pt}
   \begin{aligned}
    & UCW_j = e^{ucw}_{j,1} \odot e^{ucw}_{j,2} \odot ... \odot e^{ucw}_{j,k},\\
    & \forall t \in [1, k], t \text{ corresponds to the family of } \mathcal{S}j \\
    &  e^{ucw}_{j,1} \odot ... \odot e^{ucw}_{j,k} \in V_{UCW}\\
    & e^{ucw}_{j,1} \odot ... \odot e^{ucw}_{j,k+1} \notin V_{UCW} 
    \label{UCSequence}
\end{aligned} 
\end{equation}
where $\odot$ represents the actual symbol merging operation, which combines two symbols into one symbol and treats it as a single token in subsequent inputs, rather than representing multiple tokens that are concatenated together like in CP. $freq(\cdot)$ represents a frequency counting function, $k$ is a variable length for a UCW.

\begin{algorithm}[h]
\caption{Unsupervised Compoundation Construct Algorithm}
\label{alg:BPE}
\begin{algorithmic}[1]
\REQUIRE Music event sequence corpus $\mathcal{C}$, and the UCW's vocab size $V_{UCW}$.
\STATE $\mathcal{C}_{UCW} \gets $ Group the events into families from $\mathcal{C}$
\STATE $\mathcal{V}_{UCW} \gets \mathcal{V}_{REMI}  $ 
\WHILE {$\mathcal{V}_{UCW}.length < V_{UCW}$}
\FOR {$UCW$ in $\mathcal{C}_{UCW}$}
\FOR {$i \gets 1$ to $UCW.length -1$}
\STATE $span \gets (UCW[i],UCW[i+1])$
\STATE $freq[span] \gets freq[span]+1$
\ENDFOR
\ENDFOR
\STATE $p \gets$ the event pair with highest frequency in $freq$
\STATE $\mathcal{V}_{UCW} \gets \mathcal{V}_{UCW} \cup {p}$
\STATE $\mathcal{C}_{UCW} \gets$ replace all $p$ in $\mathcal{C}_{UCW}$ with a new symbol
\ENDWHILE
\FOR{each sequence $s$ in $\mathcal{C}$}
    \STATE Initialize an empty Set $\mathcal{S}$
    \FOR{$i = 1$ to $s.length$}
    \IF{$(s[i], s[i+1])$ not in $\mathcal{S}$}
        \STATE Add the pair $(s[i], s[i+1])$ in $\mathcal{S}$
    \ENDIF
    \ENDFOR
    \WHILE{$\mathcal{S}$ is not empty}
        \STATE Remove the pair $(s[m], s[m+1])$ from $\mathcal{S}$
        \IF{the pair $(s[m], s[m+1])$ appears in $\mathcal{V}_{UCW}$}
            \STATE Replace all $(s[m], s[m+1])$ with the new symbol $p$
            \STATE add $(s[m-1], s[n])$ and $(s[n], s[m+2])$ in $\mathcal{S}$
        \ENDIF
    \ENDWHILE
    \STATE  Append the segmented sequence $s$ to the segmented corpus $\mathcal{C'}$
\ENDFOR
\ENSURE  Corpus of UCW $\mathcal{C'}$, UCW's vocabulary $\mathcal{V}_{UCW}$
\end{algorithmic}
\end{algorithm}

Using Algorithm \ref{alg:BPE}, we derive the UCW vocabulary from our music event corpus. A comprehensive construction example can be found in Appendix A.  Through this, we've effectively reduced the sequence length than REMI while retaining the musical structure, demonstrating the power of unsupervised compoundation in symbolic music representation. As noted by \cite{hsiao2021compound}, CP sequences are typically 30\%-50\% the length of REMI sequences. With UCW, sequence length hinges on the BPE vocabulary size; for instance, a size of 1000 results in sequences 165\% the length of CP and 70\%-80\% of REMI. It is worth noting that the $V_N$ here is artificially set. Importantly, UCW, by effectively grouping neighboring elements, captures robust correlations, yielding semantically richer tokens that enhance dependency detection in our model. Similar to the BPE algorithm in NLP, it strikes a balance between REMI and CP.

\subsection{N-gram Transformer Encoder}

After constructing the UCW input sequence, the next step is to utilize an encoder for feature extraction that can effectively analyze and comprehend the music. In this section, we propose the use of an N-gram Transformer encoder, which is first pre-trained on a large-scale unlabeled UCW sequence for self-supervised learning, and then fine-tuned for downstream tasks related to symbolic music understanding.

\begin{figure*}[t]
  \centering
  \includegraphics[width=0.9\textwidth]{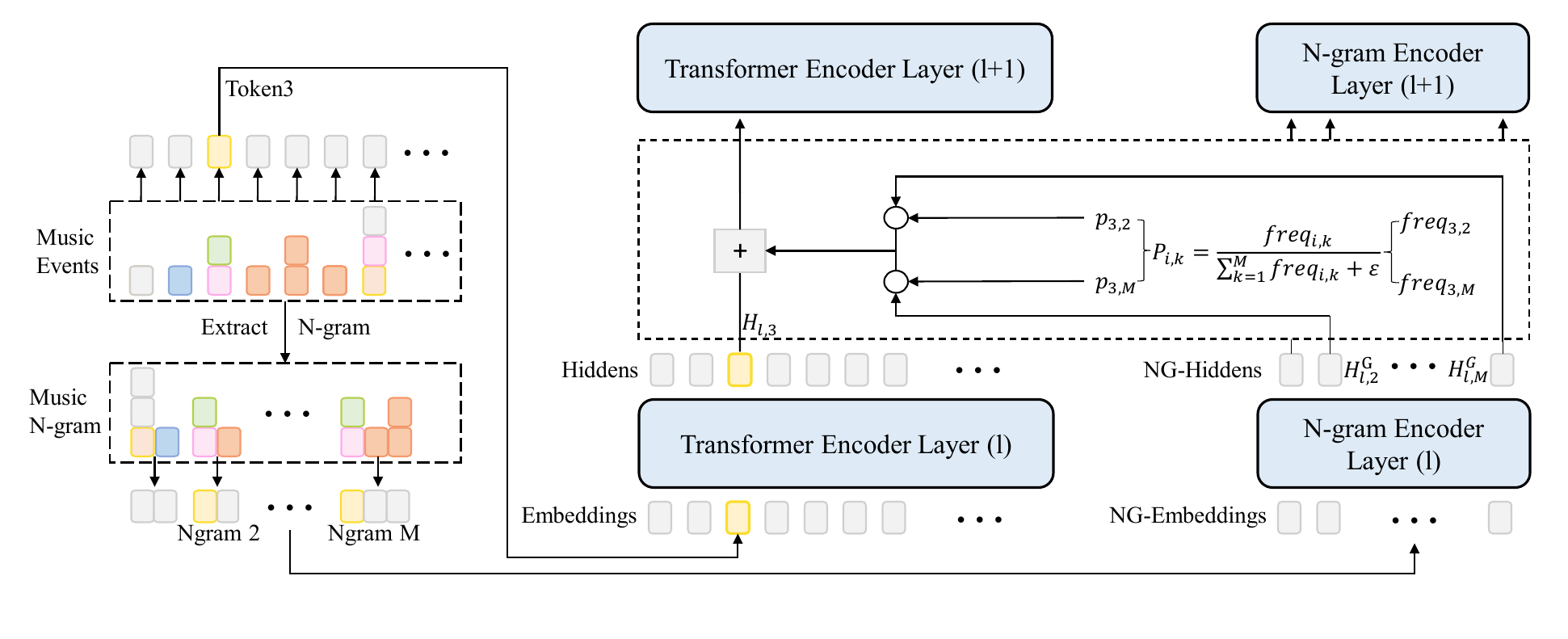}
  \caption{The overall architecture of our proposed N-gram Transformer Encoder. It only displays part of the layers of the model. In reality, after multiple layers of computation, the left Transformer Encoder will access different layer structures to accomplish specific pre-training or downstream tasks, such as Masked Language Modeling (MLM), and Sequence Classification.}
  \label{fig:NTE}
\end{figure*}

Our implementation of the N-Gram Transformer encoder is based on CP Transformer, which is a recent development in this field. However, using each CP as a single token can lead to a large vocabulary size, which may negatively impact the model's performance. To avoid this issue, CP transformers have adopted an expansion-compression approach presented in \cite{rae2019compressive}. This involves treating every element $e^{cp}_{{j,k}}$ in a CP as a separate token and embedding concatenation to integrate music elements.

In contrast, our approach views each UCW token as a locally semantically complete unit. We then transform these UCW tokens into input embeddings:
\begin{equation}
\begin{aligned}
X_{j} &= \textsc{Emb}(UCW_{j}) \\
\overrightarrow{X_j} &= X_j + \textsc{PosEmb}(j)
\end{aligned}
\label{UCE}
\end{equation}
where \textsc{Emb} converting each UCW token into a dense vector representation; \textsc{PosEmb} utilize relative position encoding, assigns vectors based on the relative distances between tokens in the sequence. $\overrightarrow{X_j}$ represents the final input of the $j$-th UCW into the Transformer layers.

Our N-gram Transformer encoder, visualized in Figure \ref{fig:NTE}, augments attentional encoding by harnessing N-gram relationships between tokens. The standard Transformer encoding is formulated as:
\begin{equation}
    H_l = \textsc{FFN}(Softmax(\frac{QK}{\sqrt{d_k}})V + H_{l-1}),
\end{equation}
where $H_{l-1}$ and $H_l$ represent the output of $l-1$-th and $l$-th layer, respectively. $Q = W^Q H_{l-1}$, $K = W^K H_{l-1}$, $V = W^V H_{l-1}$, and $H_0 = \overrightarrow{X}$. $d_k$ is the dimension of the key vectors, usually set as the hidden layer dimension divided by the number of attention heads.

To harness N-gram information in both pre-training and fine-tuning, we extract N-grams from the UCW sequence corpus, forming an N-gram vocabulary, denoted as $V_N$.
Initially, we obtain the frequency of each N-gram occurrence in the corpus and remove low-frequency N-grams to reduce the size of $V_N$. For each input UCW sequence, we extract all N-grams and select a subset of them contained within the input. We then match these extracted N-grams with the previously obtained N-gram vocabulary $V_N$, forming a sequence $G = \{g_1,g_2,...,g_M\}$, where $M$ represents the maximum number of N-grams that can be included in a single input sequence. 
These N-grams are ranked by their frequency in the corpus, ensuring top N-grams in the input are more frequent.
Similar to encoding sequences of music elements, input sequences of N-grams also need to be converted into embeddings for the encoder. We encode each N-gram token similarly as in the token Transformers encoder. Specifically,
\begin{equation}
\begin{aligned}
    &X^G_j = \textsc{Emb}^{NG}(g_j), \overrightarrow{X^G_j} = X^G_j + \textsc{PosEmb}(j)  \\
    &H^G_l = \textsc{FFN}(Softmax(\frac{Q^GK^G}{\sqrt{d_k}})V^G + H^G_{l-1}),
\end{aligned}
\label{NgramE}
\end{equation}
where $Q^G = W^Q H^G_{l-1}$, $K^G = W^K H^G_{l-1}$, $V^G = W^V H^G_{l-1}$, and $H^G_0 = \overrightarrow{X^G}$; $\textsc{Emb}^{NG}$ converting each N-gram token into a dense vector representation; 

Our current approach encodes N-Gram token sequences, aiming to integrate N-Gram information into music sequence encoding. We introduce the N-gram Position Matrix Injection (NPMI) method. With NPMI, we create a matrix called N-gram position matrix ($P$), which is an $M \times N$ matrix that records the positions and frequencies of each N-gram extracted from the input sequence. This matrix is constructed using the previously extracted N-gram vocabulary and the corresponding N-grams from each input sequence.
The N-gram position matrix captures the alignment between N-Grams and their original music sequence, which forms the core of the N-gram Transformer encoder that enhances the model's understanding capabilities using N-gram information. Specifically, based on the position of each N-gram in the sequence, we assign values to the matrix as:
\begin{equation}
    {P_{i,j} = \begin{cases}
                freq(g_j), & UCW_i \in g_j \\
                0, & UCW_i \notin g_j \\
                \end{cases}}
    \label{npm}
\end{equation}
where $UCW_i$ denotes the i-th music token, $g_j$ denotes the j-th n-gram token, and $freq(g_j)$ denotes the frequency of $g_j$ in $V_N$.

To determine the relative frequencies of N-grams, we normalize by dividing each N-gram's frequency by the total frequency of all N-grams at the same position. This normalization captures the N-gram distribution:
\begin{equation}
    P_{i,j} = \frac{P_{i,j}}{\sum_{k=0}^{M} P_{i,k}+\varepsilon}
    \label{normalization}
\end{equation}
where $\varepsilon$ is a small constant ($1\times 10^{-10}$) to prevent division by zero.

After converting the tokens of both the music input sequences and the N-gram sequences into embeddings and constructing the N-gram position matrix $P$, we combine the representations of the music tokens and their corresponding N-grams.  For each layer in the Transformer, the music token representation $H_l$ is updated by adding the corresponding N-gram representations $H^G_l$ :
\begin{equation}
    H_{l,i} = H_{l,i} + \sum_{t \in A} H^G_{l,t} \times P_{i,t}
    \label{ngramconbineforcharacter}
\end{equation}
where $A$ is the set of indexes of N-gram tokens that correspond to the music token $UCW_i$, $H^G_{l,t}$ indicate the representation of $t$-th N-Gram in $l$-th layer. Notably, our N-Gram encoding has fewer layers than the music token encoding. Thus, we only add representations for matching layers and skip addition for higher layers.

In fact, for the sake of computational efficiency, we have implemented this injection as:
\begin{equation}
\label{ngramconbineforlayer}
    H_l = H_l + P \times H^G_l
\end{equation}

\begin{table*}[htbp]
  \centering
  \small
  \caption{The testing accuracy (in \%) of various tokenization methods and models on 6 different downstream tasks.
  Symbol (*) indicates that the corresponding experimental results were replicated by us.}\label{tab:mainresult}
  \vspace{-3mm}
  \begin{tabular}{cccccccc}
  \toprule
  \multirow{2}{*}{\bf Model} & \multirow{2}{*}{\bf Token} & \multicolumn{4}{c}{\bf Sequence Classification} & \multicolumn{2}{c}{\bf Token Classification}  \\
  \cmidrule(lr){3-6} \cmidrule(lr){7-8} &  & Composer & Emotion & Genre & Dance & Velocity & Melody \\
    \midrule
    {MidiGPT \cite{ferreira2020computer} }  & CP  & -   & 61.88   & -     & -          & -  & -  \\
    \midrule
    \multirow{2}{*}{RNN \cite{chou2021midibert} }  & REMI  & 51.97   & 53.46   & -     & -          & 44.56   & 89.96  \\
          & CP-4  & 60.32   & 54.13   & -     & -          & 43.77   & 88.66   \\
    \midrule
    
    \multirow{2}{*}{OM-MIDI-Bert \cite{liu2022optimized} }  & REMI  & 82.41   & 75.58   & -     & -          & 50.82   & 92.01  \\
          & CP-4  & 75.40   & 68.81   & -     & -          & 53.42   & \textbf{97.87}   \\
    \midrule
    \multirow{2}{*}{RoAR \cite{li2022finegrained} }  & REMI  & 78.48   & 73.73   & -     & -          & 51.40   & 91.84  \\
          & CP-4  & 80.95   & 76.15   & -     & -          & 53.73   & 97.59   \\
    \midrule
    \multirow{3}{*}{MIDI-Bert \cite{chou2021midibert}} & REMI & 67.19   & 67.74   & -     & -          & 49.02   & 90.97  \\
          & CP-4  &    78.57     &    67.89     &   50.49(*)    &    57.02(*)      &    51.63     &   96.37  \\
          & CP-7  & 84.16(*)   & 72.72(*)   & 48.65(*)   &    46.89(*)     & -& -  \\
    \midrule
    \multirow{2}{*}{NG-Midiformer} & UCW-4  & 90.63        &  73.44       &  56.67     &  55.05       &   \textbf{54.13}  & 92.31 \\
          & UCW-7  & \textbf{90.63}   & \textbf{81.25}   & \textbf{65.87} &  \textbf{ 63.19}     & -       & -    \\
    \bottomrule
    \end{tabular}%
    \vspace{-3mm}
\end{table*}%

\section{Experiments}

\subsection{Setup and Downstream Tasks}

Our model undergoes a two-stage training: pre-training on a symbolic music dataset, and fine-tuning for six specific downstream tasks. Details on the dataset can be found in Appendix \ref{app:dataset}. 
We set the UCW vocabulary size at 1000, transforming MIDI music files into UCW sequences, and the resultant sequence length is just 165\% of the CP sequence.
For our N-gram approach, we extracted an N-gram vocabulary $V_N$ from the pre-trained corpus. Each N-gram was indexed and its frequency recorded. We chose an N-value of 4, excluding N-grams with frequencies below 200.

We used the same hyper-parameters as the MIDI-Bert model \cite{chou2021midibert}, which has a 12-layer structure with 12 self-attention heads, and a hidden layer size of 768 for each self-attention layer. We set a sequence length of 512 for both training stages. The N-gram encoder in our model has a 6-layer structure, with each UC sequence corresponding to an N-gram sequence of length 128. Pre-training took 44 hours (about 128k steps) on 4 NVIDIA GeForce RTX 3060 GPUs, using the AdamW optimizer and a learning rate that warmed up over the initial 1k steps. Our pre-training employed Masked Language Modeling (MLM), masking 15\% of input tokens for prediction.
Notably, MIDI-Bert \cite{chou2021midibert} and CP Transformer \cite{hsiao2021compound} differ in constructing CP sequences. While MIDI-Bert uses pitch, duration, sub-beats, and bars (CP-4), CP Transformer incorporates all seven musical elements (pitch, duration, velocity, bar, beat ,chord and tempo) with padding (CP-7). For a fair comparison, we experimented with both settings, creating UCWs (UCW-4, UCW-7) corresponding to CP-4 and CP-7 based on input music events.

We assessed our model across two main categories of tasks. For sequence classification, we focused on Composer (Com), Emotion (Emo), Genre (Gen), and Dance (Dan) classification. Meanwhile, Melody extraction (Mel) and Velocity classification (Vel) were our primary token classification tasks. A detailed introduction for these six downstream tasks and their corresponding datasets can be found in Appendix \ref{app:dataset}.

For all tasks, we fine-tuned our pre-trained model for up to 15 epochs, maintaining a consistent sequence length of 512 CP tokens. Given that a CP might encompass multiple UCWs, we input only the initial 512 UCWs during fine-tuning, which might limit the musical input in our approach. For token-level tasks, we segmented the sequence into 512 UCWs for each token's classification. However, this UCW sequence retains approximately 60\% of the CP data, potentially putting our model at a disadvantage.

Throughout both pre-training and fine-tuning, we designated 90\% of each task's dataset for training and the remaining 10\% for validation.

\subsection{Main Analysis}

Table \ref{tab:mainresult} contrasts the NG-Midiformer's performance with the baseline MIDI-Bert model across all tasks. We aligned our comparison with the CP-4 and CP-7 datasets, leading to our UCW sequences UCW-4 and UCW-7. While CP-4 captures basic musical elements, CP-7 provides a comprehensive view of piano track data. However, CP-7 and UCW-7 aren't apt for velocity and melody classification due to potential data leakage.

For sequence-level classification tasks, we standardized input music sequences to match the length of 512 CP tokens to ensure a fair comparison. In our method, only the initial 60\% of the sequence is utilized during inference. 

Yet, our NG-Midiformer, leveraging unsupervised compoundation and N-gram Transformer encoding, consistently outperformed both MIDI-Bert and RNN models in nearly all tasks. Specifically, we observed performance boosts of +8.22\%, +5.67\%, +15.08\%, +6.17\%, and +0.4\%, setting new state-of-the-art results in five tasks, including composer, emotion, genre, dance classification, and velocity prediction.

These outcomes underscore the value of N-gram information in enhancing symbolic music understanding. The fusion of UCW and N-Gram Transformer encoding surpasses the CP or REMI tokenization methods used by MIDI-Bert or RNN \cite{chou2021midibert}. Notably, UCW-7 demonstrated superior performance over UCW-4, indicating that richer information aids the model's musical comprehension.

While our model excelled in many areas, it fell short in Melody classification. This limitation might stem from using only about 60\% of the available information in UCW sequences compared to CP. This might exclude crucial melodic information present in the latter part of sequences, which can be especially important for melody classification. Additionally, our labeling method assigns identical labels to multiple UCWs within the same CP group. This means the model faces increased inference demands in token-level classification tasks, potentially affecting melody classification performance.

\begin{table}[t]
  \centering
  \small
  \setlength{\tabcolsep}{2pt}
    \caption{The testing accuracy (in \%) of various tokenization methods on 6 different downstream tasks with MIDI-Bert.
    }
  \begin{tabular}{cccccccc}
    \hline
    \multirow{2}{*}{\bf Model} & \multirow{2}{*}{\bf Token} & \multicolumn{4}{c}{Sequence Classification} & \multicolumn{2}{c}{Token Classification}  \\
    \cmidrule{3-8} & & Com & Emo & Gen & Dan&Vel&Mel \\
    \hline
    \multirow{5}{*}{MIDI-Bert} & REMI  & 67.19    & 67.74   &      - &       -     & 49.02    & 90.97  \\
              & CP-4  & 78.57    & 67.89   &  \textbf{50.49}     & \textbf{57.02 }     & 51.63    & \textbf{96.37} \\
              & CP-7  & \textbf{84.16}    & \textbf{72.72}   & 48.65 &  46.89          & -        & -  \\
              & UCW-4  &   66.61       &   63.75      & 40.34      &   55.22         &   \textbf{54.87}       &   92.26    \\
              & UCW-7  & 76.30    & 72.17   & 49.56 & 47.40      & -        & -   \\
    \hline
  \end{tabular}%
  \label{tab:onlyUC}%
\end{table}%

\begin{table}[t]
  \centering
  \small
  \setlength{\tabcolsep}{1pt}
  \caption{The testing accuracy (in \%) of various tokenization methods and models on 6 different downstream tasks.
  }
  \begin{tabular}{cccccccc}
    \hline
    \multirow{2}{*}{\bf Model} & \multirow{2}{*}{\bf Token} & \multicolumn{4}{c}{Sequence Classification} & \multicolumn{2}{c}{Token Classification}  \\
    \cmidrule{3-8} & & Com & Emo & Gen & Dan&Vel&Mel \\
    \hline
    \multirow{4}{*}{MIDI-Bert} & CP-4  & 78.57    & 67.89   & 50.49     & 57.02          & 51.63    & 96.37\\
              & CP-7  & 84.16    & 72.72   & 48.65 &    46.89        & -       & -    \\
              & UCW-4  & 66.61      &   63.75      &  40.34     &  55.22          &  54.87        &  96.26    \\
              & UCW-7  & 76.30    & 72.17   & 49.56 &  47.40          & -        & -   \\
  \midrule
    \multirow{4}{*}{NG-Midiformer}     & CP-4  &      83.33   &   65.87      & 53.89      &  61.30          &   \textbf{59.82}       &  \textbf{96.92}    \\
              & CP-7  & 85.94    & 70.63   &  65.63     &   47.62         & -        & -     \\
              & UCW-4  &   90.63       &  73.44       & 56.67      &  55.05  &  54.13      &  92.31  \\
              & UCW-7  & \textbf{90.63}    & \textbf{81.25}   & \textbf{65.87} &  \textbf{63.19}       & -        & -   \\
    \hline
  \end{tabular}%
  \label{tab:onlyngram}%
\end{table}%


\subsection{Ablation Study}

In this section, we will examine the effects of unsupervised compoundation, N-Gram Transformer encoding and pre-training with N-gram respectively.

\subsubsection{Effect of Unsupervised Compoundation}

Unsupervised compoundation, a tokenization method lying between REMI and CP, was evaluated for its impact on symbolic music understanding. Using the same structures as MIDI-Bert, we compared five tokenization methods on downstream tasks. As Table \ref{tab:onlyUC} reveals, CP-4 and CP-7 outperformed others. The diminished performance with UCW-4 or UCW-7 suggests their heavy reliance on N-gram encoding. MIDI-BERT might expect a different granularity level than what UCW provides, leading to inefficiencies in capturing the nuances of music sequences. For a fair comparison with CP, we standardized sequence lengths to 512 CPs. However, this meant only the first 512 UCW tokens were utilized, potentially affecting performance. This underscores the significance of N-Gram feature injection for model enhancement.

\begin{table}[t]
  \centering
  \small
  
  \setlength{\tabcolsep}{1pt}
    \caption{The testing accuracy (in \%) of on different methods for initializing downstream tasks' models.}
  \begin{tabular}{lccccc}
    \toprule
    \multirow{2}{*}{\bf Model} & \multirow{2}{*}{\bf Token} & \multicolumn{4}{c}{Sequence Classification}  \\
    \cmidrule{3-6} & & Composer & Emotion & Genre & Dance \\
    \midrule
    Midi-Bert  & UCW-7 & 76.30 & 72.17& 49.56 & 47.40 \\
    \midrule
    NG-Midiformer  & UCW-7 & \textbf{90.63} & \textbf{81.25} & \textbf{65.87} & \textbf{63.19}  \\
    \quad w/o N-Gram Pre-training & UCW-7 & 58.73 & 68.97 &31.03  & 52.63 \\
   \hline
  \end{tabular}%
  \label{tab:pretrain}%
\end{table}%

\subsubsection{Effect of N-gram Transformer Encoding}

We evaluated the N-gram Transformer Encoder against MIDI-Bert using four tokenization methods, as shown in Table \ref{tab:onlyngram}. The results highlight the N-gram's advantage in music modeling, especially when combined with UCW tokenization in our NG-Midiformer. However, the two CP methods didn't benefit from N-gram in emotion classification, possibly due to lost correlations during music event merging, and N-gram encoding might not be sufficient to recover them, especially if they arise from longer-range dependencies.

\begin{figure}
\centering
    \begin{tikzpicture}[scale=0.7]
    \begin{axis}[
        width=0.45\textwidth,
        xlabel={Step (k)},
        ylabel={Pre-Training Loss},
        xmin=0, xmax=15,
        ymin=0, ymax=6,
        xtick={0,1,2,3,4,5,6,7,8,9,10,11,12,13,14,15,16},
        ytick={0,1,2,3,4,5,6},
        legend pos=north east,
        grid=major,
    ]
    
    \addplot[
        color=blue,
        mark=square,
        ]
        coordinates {
            (0.5,5.50)(1,5.37)(2,5.27)(4,5.18)(6,5.01)(7,4.37)(8,3.53)(10,3.15)(12,2.96)(15,2.79)
        };
        \addlegendentry{Midi-Bert}
        
    \addplot[
        color=red,
        mark=*,
        ]
        coordinates {
           (0.5,2.48)(1,1.564)(2,1.31)(4,1.01)(6,0.98)(7,0.95)(8,0.92)(10,0.87)(12,0.84)(15,0.81)
        };
        \addlegendentry{NG-Midiformer}
    \end{axis}
    \end{tikzpicture}
        \caption{The convergence rate of the Bert model and our model (Taking UC-4 as an example ).}
        \label{fig:speed}
\end{figure}

\subsubsection{Effect of pre-training with N-gram}

To evaluate the effectiveness of using N-gram information for pre-training, we conducted experiments to assess the model's performance on downstream tasks. We excluded the N-gram pre-training stage  and present the results of these experiments in Table \ref{tab:pretrain}.
When comparing Midi-Bert with our NG-Midiformer, we found that our NG-Midiformer results far exceeded Midi-Bert under the same input, indicating the importance of N-Gram feature injection.

However, without its pre-training, the N-Gram Transformer's performance, relying on mere random parameter updates during fine-tuning, fell below Midi-Bert, underscoring pre-training's crucial role. Additionally, with consistent hyperparameters and pre-training datasets, our N-gram Transformer encoder converged more rapidly than Midi-Bert, indicating our model's heightened efficiency in interpreting symbolic music sequences (Figure \ref{fig:speed}).

\section{Conclusion}

In this paper, we introduce the NG-Midiformer model, which proposes an unsupervised compoundation and N-gram feature injection approach. Due to the strong correlation between neighbouring elements in music, N-gram is particularly suitable for symbolic music understanding and enhances the model's ability to understand symbolic music. Our method first reconstructs the input token sequence through unsupervised methods, then pre-trains the model on a large-scale corpus using N-gram information to enhance music understanding, and finally fine-tunes on different downstream tasks to complete specific tasks. The experimental results and analysis of our method demonstrate the effectiveness of our approach. In the future, we will apply our proposed method to music generation tasks to generate various styles and genres of music.

\bibliography{aaai24}
\clearpage
\appendix
\section{APPENDIX}

\subsection{Datasets}

Table \ref{tab:dataset-usage} presents the dataset we used in this study.
In the pre-training stage, we extracted and transformed the four datasets listed in the table, and ultimately obtained 24,979 UCW sequences that corresponded to the music midis. In the previous work, symbolic music datasets such as GIantMIDI Piano \cite{kong2020giantmidi}, Maestro \cite{hawthorne2018enabling}, and Pop1K7 \cite{hsiao2021compound} are often utilized for pre-training. Additionally, ADL Piano \cite{ferreira_aiide_2020}, a widely used large-scale music dataset, is also incorporated for pre-training. By integrating these diverse symbolic music datasets, we aim to improve the model's generalization capabilities and expand its knowledge base. 
In the fine-tuning stage, we selected six different music classification tasks with 5 datasets to verify the performance of the model.

The GTZAN dataset \cite{sturm2013gtzan} here is in wav format. We first convert it into MIDI files using the piano transcription system~\cite{kong2021high}, which is a widely used neural network for transcribing music audios into MIDI format, and then convert them into the corresponding UCW sequences. 
We also collected the Nottingham dataset~\cite{Nottingham}, a classic folk music collection, from an online source. To evaluate our model's performance, we utilized the MIDI version of this dataset specifically for the task of dance music classification. 
GTZAN dataset and Nottingham dataset are used for genre classification and dance classification, respectively.
\label{app:dataset}
\begin{table}[hp]
    \centering
    \small
     \caption{Summary of dataset usage.}
    \label{tab:dataset-usage}%
    \begin{tabular}{{p{0.20\textwidth}p{0.14\textwidth}p{0.03\textwidth}p{0.03\textwidth}}}
    \toprule
    Dataset & Usage & Pieces & Hours \\
    \midrule
    Pop1K7\cite{hsiao2021compound} & Pre-training & 1,747 & 108.8 \\
    ADL Piano\cite{ferreira_aiide_2020}   & Pre-training & 11086&1775 \\
    GIantMIDI-Piano\cite{kong2020giantmidi} & Pre-training & 10855 & 1237\\
    Maestro\cite{hawthorne2018enabling}   & Pre-training & 529& 84\\
    \multirow{2}{*}{POP909\cite{wang2020pop909}} & Melody Extraction &\multirow{2}{*}{865} & \multirow{2}{*}{59.7} \\
    & Velocity Prediction &&\\
    Nottingham & Dance Classification & 1034 & 19.42\\
    Pianist8\cite{joann8512_2021_5089279} & Composer Classification & 411 & 31.9\\
    EMOPIA\cite{EMOPIA} & Emotion Classification & 1,078 & 12.0\\
    GTZAN\cite{sturm2013gtzan} & Genre Classification & 1000 &8.3 \\
    \bottomrule
    \end{tabular}%
\end{table}%
In line with our previous work, we utilized the POP909~\cite{wang2020pop909} dataset for tasks related to Melody Extraction and Velocity Prediction. Additionally, we employed the EMOPIA dataset~\cite{EMOPIA} and Pianist8 dataset~\cite{joann8512_2021_5089279} for sentiment classification and composer classification tasks, respectively.

Velocity classification and melody extraction are tasks of token classification, whereas genre, emotion, composer and dance classification are tasks of sequence classification.

\subsection{Constructing Unsupervised Compoundation Word}
Consider a short sequence: [``Bar", ``Beat\_0", ``Tempo\_119", ``G\_M", ``Pitch\_71", ``Duration\_1080", ``Velocity\_90", ``Pitch\_69", ``Duration\_1560", ``Velocity\_90", ``Bar", ``D\_7``, ``Pitch\_71", ``Duration\_1080", ``Velocity\_88", ``Pitch\_73", ``Duration\_1560", ``Velocity\_90"], where G\_M and D\_7 are two different chords. Firstly, based on the concept of family, we group music events and obtain the following sequence: [``Bar", ``Beat\_0 Tempo\_119 G\_M", ``Pitch\_71 Duration\_1080 Velocity\_90", ``Pitch\_69 Duration\_1560 Velocity\_90", ``Bar D\_7``, ``Pitch\_71 Duration\_1080 Velocity\_88", ``Pitch\_73 Duration\_1560 Velocity\_90"].The algorithm\ref{alg:BPE} identifies the most frequent event pairs with a token, such as (``Pitch\_71", ``Duration\_1080"), and merges them as a new token. After the first iteration, the sequence becomes [``Bar", ``Beat\_0 Tempo\_119 G\_M", ``Pitch\_71 Duration\_1080", ``Velocity\_90", ``Pitch\_69 Duration\_1560 Velocity\_90", ``Bar D\_7", ``Pitch\_71 Duration\_1080", ``Velocity\_88", ``Pitch\_73 Duration\_1560 Velocity\_90"]. Next, we identifies the next most frequent pairs (``Duration\_1560", ``Velocity\_90"), and the sequences becomes [``Bar", ``Beat\_0 Tempo\_119 G\_M", ``Pitch\_71 Duration\_1080", ``Velocity\_90", ``Pitch\_69", ``Duration\_1560 Velocity\_90", ``Bar D\_7", ``Pitch\_71 Duration\_1080", ``Velocity\_88", ``Pitch\_73", ``Duration\_1560 Velocity\_90"]. From this, we can derive the UCW sequence and its associated vocabulary for this sequence.

\subsection{Recall and Precision Results}
In addition to achieving uniform classification accuracy, we also assessed the recall and precision metrics for our UCW-7 and UCW-4 models in the context of four downstream sequence classification tasks on our dataset, for reference purposes.
\begin{table}[h!]
\centering
\small 
\setlength\tabcolsep{3pt} 
\begin{tabular}{@{}ccccccccc@{}}
\hline
& \multicolumn{2}{c}{Composer} & \multicolumn{2}{c}{Emotion} & \multicolumn{2}{c}{Genre} & \multicolumn{2}{c}{Dance} \\ \cmidrule{2-9}
& P & R & P & R & P & R & P & R \\ \hline
UCW-7 & 0.84 & 0.77 & 0.76 & 0.78 & 0.67 & 0.54 & 0.39 & 0.43 \\
UCW-4 & 0.85 & 0.84 & 0.69 & 0.69 & 0.41 & 0.49 & 0.41 & 0.45 \\ \hline
\end{tabular}
\caption{The precision and recall values of NG-Midiformer on four sequence classification tasks using two types of tokens, where P represents precision and R represents recall.}
\label{PR}
\end{table}
\end{document}